\documentstyle[12pt,moriond,psfig]{article}
 
\begin{document}

\heading{THE SPECTRAL ENERGY DISTRIBUTION OF SPIRAL GALAXIES}

\author{H. R. Schmitt $^{1}$}
{$^{1}$ Space Telescope Science Institute, 3700 San Martin Drive, Baltimore, MD21218, USA}   

\begin{moriondabstract}

Spectral energy distributions (SEDs) are some of the most important
sources of information for galaxies, especially for high redshift ones.
Here we review recent work on the subject. We discuss the integrated
spectra of galaxies of different morphological and activity type, their
application to derive K-corrections and classification of high redshift
objects. We also discuss the radio to X-rays SEDs of Seyfert 2's,
Starbursts and normal galaxies, their behavior as a function of the
waveband, their bolometric fluxes and the wavebands that contribute
most to it.

\end{moriondabstract}

\section{Introduction}

With the increasing number of surveys of high redshift galaxies, it is
necessary to have a good knowledge of the integrated properties and the
stellar population of local galaxies, since this is virtually the only
information available for distant objects
\cite{Dressler},\cite{Ellis},\cite{EllisAR}. Integrated spectra of
local galaxies can be used for the spectroscopic classification of
galaxies discovered through radio, X-rays and infrared surveys, to
compute K-corrections, galaxy number counts, and to determine the
redshifts or morphological types of distant objects based on their
colors.

Similarly, the knowledge of the continuum energy output of galaxies
over a broad wavelength range is important for the understanding of the
physical processes responsible for the emission at different wavebands.
It is also a means to distinguish between objects of different activity
class, and for the determination of accurate bolometric fluxes.

\section{Integrated Spectra of Galaxies}

Some of the first and most used integrated spectra of galaxies and
K-corrections were the ones compiled by Pence (1976) and Coleman et al.
(1980) \cite{Pence},\cite{Coleman}. These were the best template
spectra available for several years. However, they were based on
inhomogeneous samples of ultraviolet and optical spectra of galaxies,
and, in some cases, were made combining spectra of galaxies and stars.

Nowadays, the best set of spatially integrated template spectra of galaxies
available in the literature is that of Kennicutt (1992)
\cite{Kennicutt}. It includes 55 galaxies of different morphological
and activity types, observed from the ground with apertures of
$\approx90^{\prime\prime}\times90^{\prime\prime}$, which in most cases
includes the entire galaxy. The only problem with these spectra is that
they span only a short wavelength range $\lambda$3650-7000\AA, and
cannot be used for the optical study of high redshift galaxies
(z$>1$).

The other set of template spectra available in the literature is that
of Kinney et al. (1996) \cite{Kinney}, which includes both ground based
and satellite ultraviolet (IUE) data of 69 galaxies (quiescent and
starburst galaxies), observed with matched apertures
(10$^{\prime\prime}\times20^{\prime\prime}$). Although these spectra
were observed with an aperture much smaller than that used by
\cite{Kennicutt}, they cover the wavelength range 1200-10000\AA\ and
can be used in the study of higher redshift galaxies (up to z$\approx$2
in the B band and even higher for redder wavebands).

\begin{figure}
\vskip -5.0cm
\psfig{figure=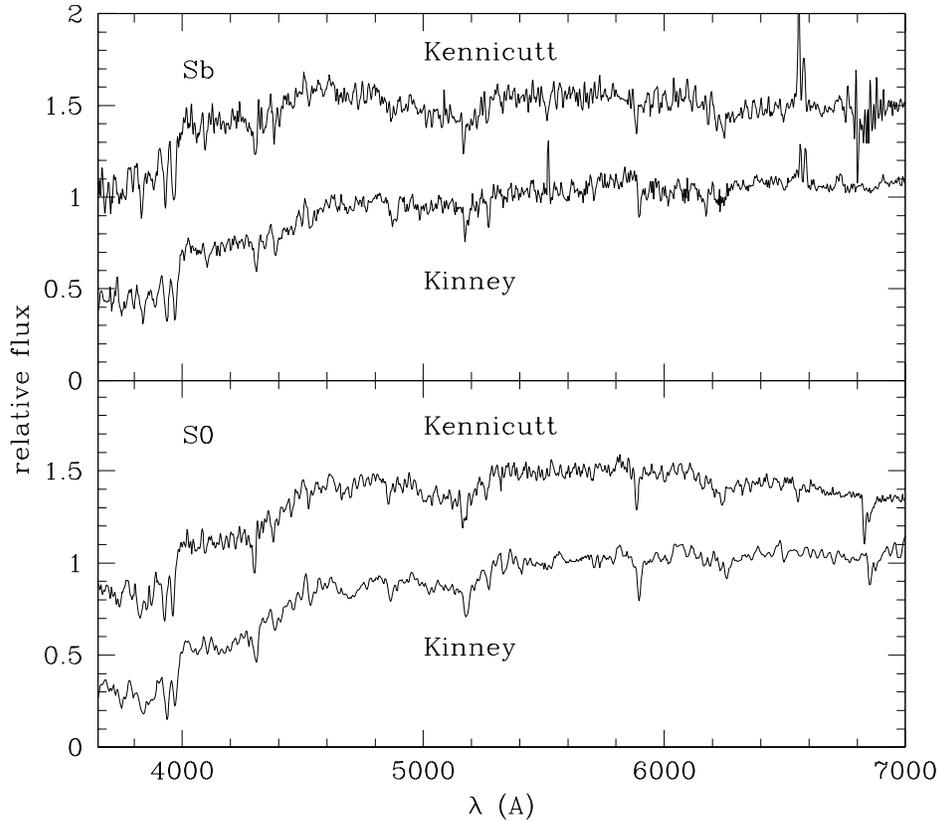,width=16cm,height=16cm}
    \caption{Comparison between Kennicutt (1992) and Kinney et al. (1996)
Sb (top) and S0 (bottom) templates. The templates were normalized to the
flux at $\lambda$5500\AA\ and displaced by a constant for clarity.}
\end{figure}
 
In Figure 1 we show the Kinney and Kennicutt S0 (bottom) and Sb (top)
templates. The comparison between the equivalent widths of metal and HI
lines, as well as continuum shapes, indicates that both sets of S0
templates are very similar and the Kinney templates do not suffer from
aperture effects. In the case of Sb galaxies, the comparison shows that
the Kinney template has an older stellar population and redder
continuum, indicating that the use of smaller apertures probably
included most of the bulge emission and missed part of the emission
from young stellar populations in the disk.

Figure 1 also illustrates the evolution of the templates characterists
as a function of the morphological type. Early type galaxies
(ellipticals, S0's and Sa's) are characterized by strong metal
absorption lines (e.g.  Ca H and K around $\lambda$3900\AA\ and the Mg
lines around $\lambda$5200\AA) and a large amplitude 4000\AA\ break,
typical of old stellar populations.  As we move to late type galaxies
(Sb's, Sc's, Sd's and irregulars), the intensity of the metal
absorption lines decreases, we start to see strong high order HI lines
in absorption (around 4000\AA), typical of young stars, the amplitude
of the 4000\AA\ break decreases, and emission lines appear (e.g.
H$\alpha$, [NII] and [OII]).

One of the applications of the templates can be seen in Figure 2, where
we show the predicted colors of galaxies of different morphological
type (in the case of quiescent galaxies) and reddening (in the case of
starburst galaxies), created by redshifting the templates, with no
evolution, and measuring their colors. These kinds of plots can
be used to determine the morphological type of galaxies, once their
redshifts and colors are known, or in the case where the morphological
types and colors are known it can be used to determine the redshifts.

\begin{figure}
\psfig{figure=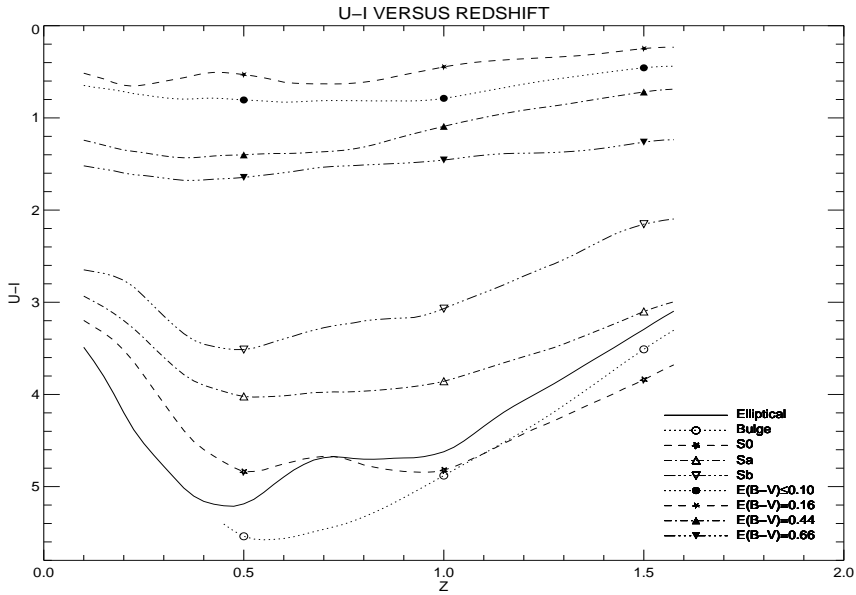,angle=90,width=12cm,height=8cm}
    \caption{U-I color versus redshift, predicted using the Kinney
templates \cite{Kinney} and assuming no evolution.}
\end{figure}

For other applications of these templates, like K corrections, number
counts of galaxies, determination of redshifts based on colors and star
formation properties of galaxies, we suggest the following references
\cite{Conol},\cite{Ellis},\cite{Kennicutt},\cite{KennicuttAR},\cite{Kinney}.

\section{Spectral Energy Distributions}

Most of the previous works in this area (e.g. \cite{Edelson},\cite{Elvis},\cite{Sanders},\cite{Wilkes}) were
concentrated on the study of the nuclear emission of high luminosity
objects, like Quasars and Seyfert 1 galaxies, objects for which it is
relatively easy to find large quantities of data on several
wavebands, and where the nuclear emission dominates over the host
galaxy flux, so aperture effects are not a problem for the analysis.

Until recently, very little had been done on the SEDs of Starbursts,
Seyfert 2's, LINER's and Normal galaxies, to study of the properties of
the entire galaxy and not only the emission from the nuclear region,
like in high luminosity AGN's. Some works like
\cite{Mas94},\cite{Mas95},\cite{Spinoglio} presented radio to X-ray
multiwavelength analysis of Seyfert 1's, Seyfert 2's, Starbursts,
Quasars and normal galaxies. Although some of these works dealt with
large samples, they only used relatively sparse data points to cover
the entire frequency range.  Also, in the case of \cite{Spinoglio}, the
authors applied corrections to include the flux of the entire galaxy
in the analysis, which is uncertain, and did not distinguish between
normal and Starburst galaxies.

Here we discuss the SEDs of normal, active and Starburst galaxies
presented by Schmitt et al. (1997) \cite{Schmitt}. These SEDs were
created using measurements collected from large databases of X-rays,
ultraviolet, infrared and optical, spanning over 10 decades in frequency
(10$^8$---10$^{18}$ Hz).

\subsection{Data and Aperture Effects}

In Figure 3 we show the individual SED's of spiral galaxies, normalized
to the flux at $\lambda$7000\AA, which will be used to illustrate 
the sources from which the data were obtained. This Figure will also
be used to explain possible aperture effects and the processes responsible
for the emission at different wavebands.

\begin{figure}
\vskip -6.5cm
\psfig{figure=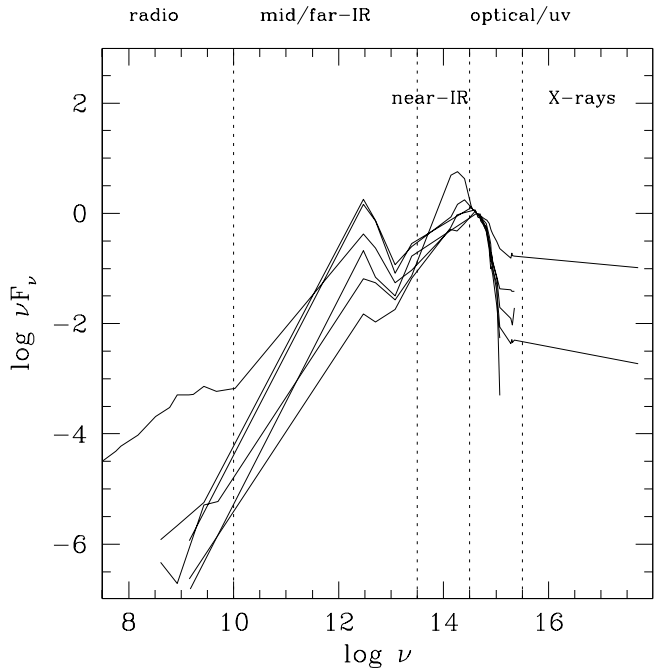,width=16cm,height=16cm}
    \caption{SEDs of normal spiral galaxies, normalized to the
flux at $\lambda$7000\AA. The figure also shows the division
between different wavebands: radio (log $\nu<$10),
mid/far-IR (11$<$log $\nu<13.5$), near-IR (13.5$<$ log $\nu<14.5$),
optical/ultraviolet (14.5$<$ log $\nu<15.5$), X-rays (log $\nu>15.5$).}
\end{figure}

The data used in the construction of these SEDs is described in
\cite{Schmitt}. It consists of ultraviolet IUE and ground-based spectra
for 59 galaxies, obtained with similar apertures
(10$^{\prime\prime}\times20^{\prime\prime}$) by
\cite{Kinney},\cite{McQuade},\cite{Storchi}, and literature near-IR
data (J,H,K and L band photometry) also obtained with similar
apertures.  For the X-ray data we used Einstein observations (0.2-4Kev)
from the literature, and the mid/far-IR data was obtained from the IRAS
catalog, both of which include emission from the entire galaxy. The
radio data were obtained from the literature, and in some cases
only data observed with large apertures could be found. One gap in
these SEDs is data in the millimeter waveband, which was available only
for 3 Seyfert 2 galaxies in our sample.

The emission in the different wavebands originates from different
processes in different types of galaxies. The X-ray originates from
thermal emission from gas heated by supernovae remnants in the case of
normal spirals and Starbursts, and mostly from cooling flows in normal
ellipticals. In Seyfert 2's, and, possibly in LINER's, it originates
from the accretion disk.  The ultraviolet emission is dominated by
young stars, can have some contribution from the AGN in the case of
Seyfert 2's and LINER's, but has very little contribution from old
stars. The visual and near-IR parts of the SEDs are dominated by the
old stellar population.  The mid/far-IR emission comes from ultraviolet
and visual radiation absorbed by dust and reradiated in this waveband.
The radio emission originates from free-free emission in HII regions
and supernova remnants in normal spirals and Starbursts, is mostly
synchrotron emission from the active nucleus in Seyfert 2's and
LINER's, while in ellipticals it can originate from free-free emission
in a cooling flow or also synchrotron emission from an active nucleus.

As said above, we tried as much as possible to use measurements
obtained with apertures similar to that of the IUE
(10$^{\prime\prime}\times20^{\prime\prime}$), which were relatively
easy to obtain for the ultraviolet, visual and near-IR wavebands.
However, in the case of X-rays, mid/far-IR, and for some of the
galaxies in radio, we had to use measurements that included the entire
galaxy. We estimate that the use of such large aperture data does not
influence the results significantly in the case of Seyfert 2's and
Starbursts, because most of the light in these objects is concentrated
in the nuclear region, but it may present some problem in the case of
the lower luminosity objects (LINER's and normal galaxies).

Figure 3 demonstrate some of the effects caused by the use of data with
large apertures to construct the SEDs of normal spirals. For X-rays,
only data for 2 normal spirals were available. The object with the
strongest emission is NGC598, which is a nearby late type spiral
galaxy. The emission in this waveband is strongly influenced by the
emission from HII regions and supernova remnants along the galaxy disk,
which makes this portion of the SED look much stronger than it actually
would be if observed with an aperture similar to that of IUE. In the
case of elliptical galaxies, the X-ray emission can be influenced by
cooling flows. A similar problem also happens in the radio waveband,
where large aperture data can include emission from HII regions and
supernova remnants in the disk of spirals, while in ellipticals it can
be strongly influenced if the galaxy contains a radio loud nucleus, or
if it is in the middle of a cooling flow. The mid/far-IR waveband can
be affected by two different effects, the amount of dust in the galaxy
and the dust temperature. Since dust absorption is more effective at
shorter wavelengths, galaxies with stronger ultraviolet emission have
stronger mid/far-IR emission (hotter temperatures) than galaxies with
similar dust content, but redder stellar population.

\subsection{X-ray to Radio SEDs}

The galaxies were separated in six groups, according to the
morphological and activity type: normal spirals (6 galaxies),
normal ellipticals (7 galaxies), LINER's (5 galaxies), Seyfert 2's (15
galaxies), Starbursts of low reddening (11 galaxies) and Starbursts of
high reddening (15 galaxies). The last two groups will be called SBL
and SBH, respectively, hereafter.  The SBL group comprises those
Starbursts with E(B-V)$<$0.4, while the SBH's are those with
E(B-V)$>$0.4 (assuming the values given by \cite{Calzetti}).

In Figure 4 we show the average SED's of these six groups of galaxies,
where the errorbars represent the standard deviations and can be used
as a guide of how similar, or different the individual SED's are
in different waveband. It can be noticed that the SEDs of normal spirals
show a considerable spread in the ultraviolet to near-IR, probably
due to their different morphological type, where some of them have
HII regions close to the nucleus, which increases the ultraviolet
flux. As discussed above,
there were X-ray data available only for two of these galaxies and one
of them is a large nearby object, which makes the X-ray SEDs very different
between the two. The emission in the mid/far-IR has a large spread,
which can be attributed to both aperture effects and the amount of
dust in the galaxies. Finally, the SEDs are very similar in the radio
waveband, with the exception of NGC598, discussed above, which was
observed with a large aperture.

The SEDs of elliptical galaxies unlike the spirals, are very similar
in the ultraviolet to near-IR range, consistent with an old, red stellar
population, and the presence of the UV turn-up. The situation changes
when we go to the mid$/$far IR and radio wavebands, where there is a
large difference between individual SEDs. The large spread in
mid$/$far-IR properties can be attributed to different amounts of dust
\cite{Goudf}, while in the radio, it can be due to cooling flows or the
existence of a radio loud nucleus. In part, the differences in the
radio emission could also be due to the different apertures through
which the observations were taken.  The X-ray fluxes of these galaxies
also show some differences, which are due to the large aperture through
which they were observed, including the contribution from sources like
X-ray binaries and the hot gaseous halo \cite{Fabbiano}, which extend
for much more than 10$^{\prime\prime}\times20^{\prime\prime}$.

\begin{figure}
\vskip -1.0cm
\psfig{figure=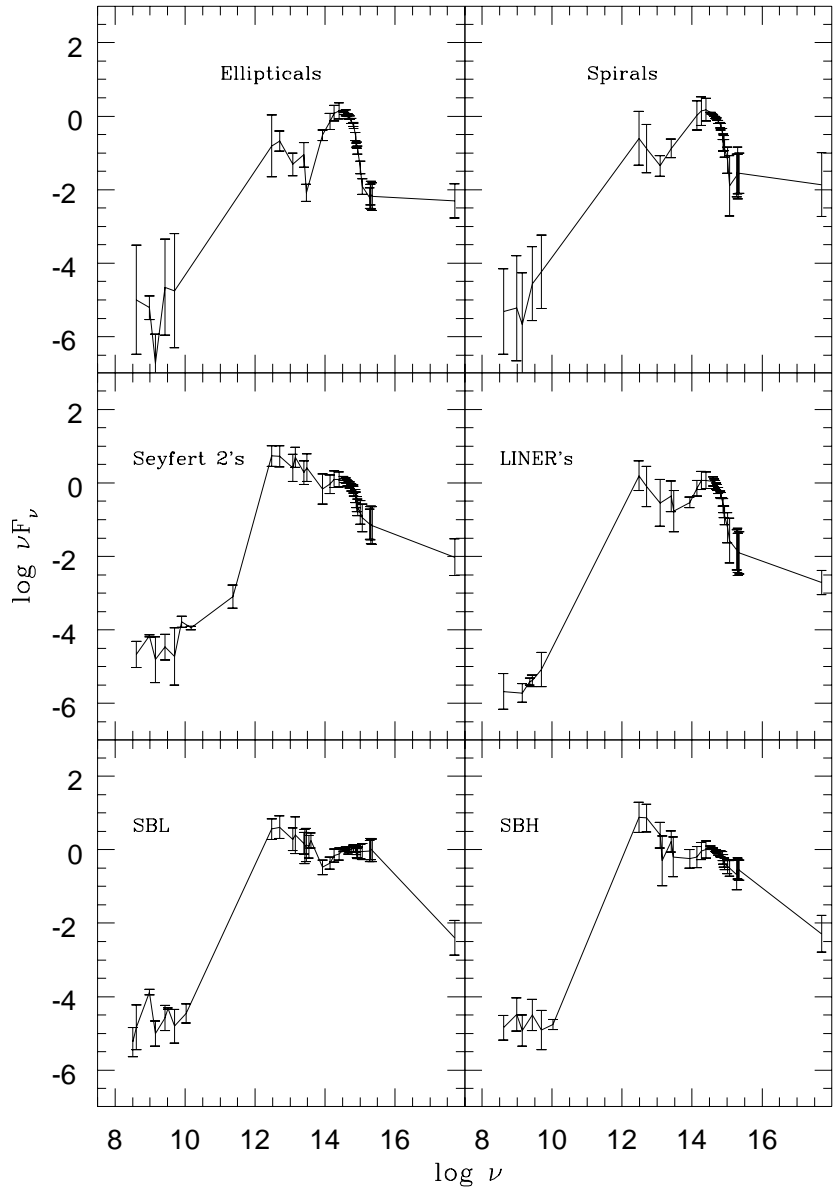,width=16cm,height=16cm}
    \caption{Average SED's of normal ellipticals (top left), normal
spirals (top right), Seyfert 2's (middle left), LINER's (middle right),
low reddening starbursts -- SBL -- (bottom left) and high reddening
starbursts -- SBH -- (bottom right). The error bars indicate the
standard deviation of the average.}
\end{figure}
 
Seyfert~2 galaxies have similar SEDs in the near-IR to radio
wavelengths.  However, when we move to the visual and ultraviolet
range, the SEDs start to show a considerable variation from object to
object, being as red as a normal galaxy or as blue as a Starburst. This
increasing blueness can be due to an increasing contribution from the
AGN continuum to the spectrum, or to the presence of circumnuclear HII
regions. Figure 4 also shows that there is a steep drop in the emission
from far-IR to the millimeter waveband (Log$\nu\approx$11.5), which
represents the end of the thermal emission from radiation reprocessed
by the circumnuclear torus, and maybe HII regions in the galaxy disk,
and the beginning of the non-thermal, synchrotron radio emission.

The LINER SEDs are similar in the radio and visual part of the
spectrum, but have some spread in mid/far IR and ultraviolet wavebands. The
mid/far IR spread can be explained using the same arguments used
above for normal galaxies, while the difference in the ultraviolet band can be
due to an increasing contribution from a population of young stars, or
the active nucleus.  The emission in the X-ray has some spread due to
the large aperture.

The SBL's and SBH's have similar SEDs over the entire energy spectrum.
The SBL's have a small spread in the ultraviolet, while for SBH's the most
noticeable spread is in the radio and far IR bands. The X-ray emission,
contrary to what is observed for the rest of the galaxies, drops
abruptly relative to the ultraviolet emission in both types of Starburst
galaxies.  The emission in the X-ray comes mostly from SNR,
concentrated in the Starburst region. When comparing the SEDs of SBL's
and SBH's (see also Figure 5), we notice that SBL's have stronger
ultraviolet emission than SBH's, while in the far-IR the opposite
happens. This difference is due to the fact that the ultraviolet
and visual radiation absorbed by dust in SBH's is reradiated in the far-IR
\cite{Cal95}.

\begin{figure}
\vskip -5.0cm
\psfig{figure=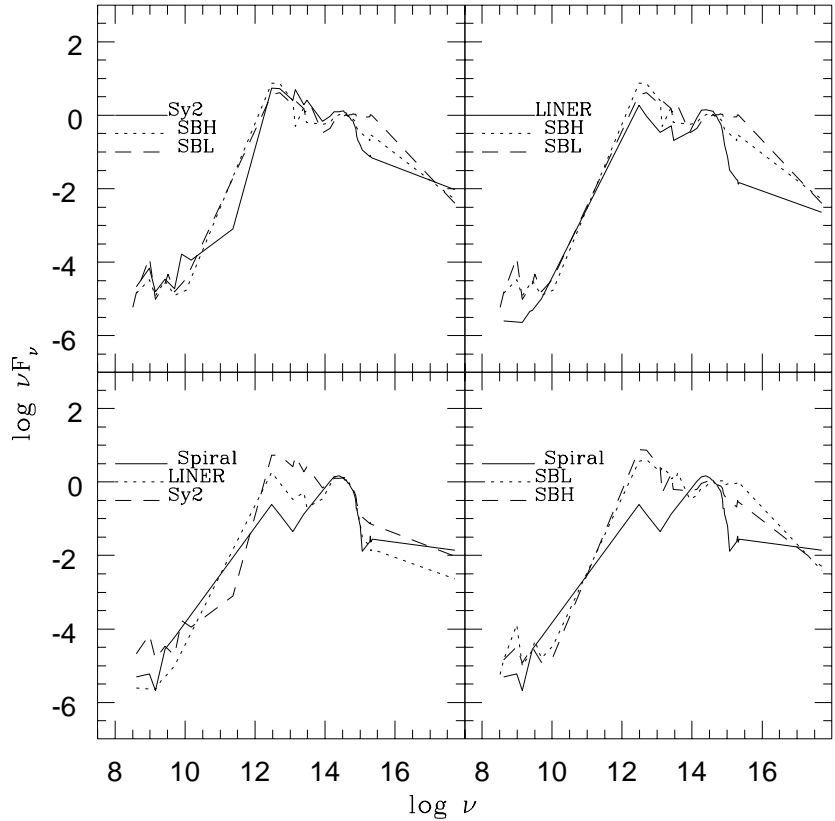,width=16cm,height=16cm}
    \caption{Comparison between the average SED of Seyfert 2, high and
low reddening Starbursts (top left); LINER's, high and low reddening
Starbursts (top right); normal spirals, LINER's and Seyfert 2's (bottom
left); and normal spirals, high and low reddening Starbursts (bottom right).}
\end{figure}

In Figure 5 we compare the SED's of galaxies with different activity
class. On the top left panel we plot the Seyfert 2's, SBL's, and SBH's,
which have similar SEDs in the radio to near-IR waveband, but start to
diverge in the visual and ultraviolet parts of the spectrum. In this
waveband the Seyfert 2's are mostly dominated by the old stellar
population and have the reddest energy distribution, probably due to
the obscuration of the AGN continuum by the torus. On the other hand,
SBH's and SBL's are increasingly bluer, and dominated by the young
stellar population. These SEDs also differ in the X-ray waveband where
Seyfert 2's are brighter. The comparison between the SEDs of LINER's,
SBH's and SBL's is made on the top right panel of Figure 5.  The only
waveband region where these SEDs are similar is from the visual to the
near-IR, where they are normalized.  The LINER's SED is systematically
fainter at all other bands.

The SEDs of LINER's, Seyfert 2's, and spirals are compared on the
bottom left panel of  Figure 5. LINER's and spirals have very similar
SEDs, only differing in the mid/far-IR and ultraviolet, where the
spirals are fainter, and in the far-IR where LINER's are brighter. The
similarity between these two SEDs is due to the fact that the nuclear
luminosity of LINER's is low, so their SED are dominated by emission
from the host galaxy.  Seyfert 2's and spirals SEDs are similar only in
the near-IR to visual waveband, where they are dominated by the old
stellar population, with the Seyfert 2's being much brighter than the
spirals in the IR and ultraviolet.  The SEDs of spirals, SBL's and
SBH's are compared on the bottom right panel of Figure 5. In this panel
we can see the difference between SEDs dominated by old (spirals) and
young stellar populations (SBH's and SBL's). The only wavelength region
where these SEDs can be considered similar is in the visual to near-IR,
again the region where they are normalized (Starbursts have some
contribution from old stars in this region). The spirals are fainter in
any other waveband.

\begin{figure}
\vskip -9.0cm
\psfig{figure=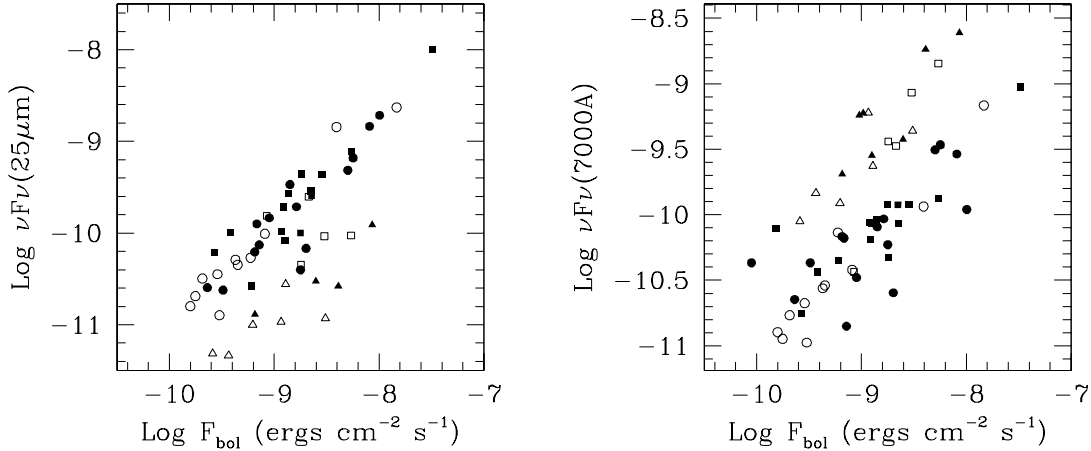,width=16cm,height=16cm}
    \caption{Relations between Bolometric flux and the flux density at
25 $\mu$m (left) and 7000\AA\ (right). Filled squares represent Seyfert 2's,
open squares LINER's, filled triangles normal ellipticals, open triangles
normal spirals, filled circles SBH's and open circles SBL's}
\end{figure}
 
\subsection{Bolometric Fluxes}

Another important point in the study of the SEDs of galaxies is their
use in the measurement of bolometric fluxes, which was done by integrating
the SEDs over the entire wavelength range. Some of the results are
presented in Figure 6, and we direct the reader to Section 8 of
\cite{Schmitt} for a better description of the results. In this Figure
we compare the bolometric flux with the flux density at 25$\mu$m (left)
and 7000\AA\ (right).  The bolometric flux of both Seyferts and
Starbursts show a good correlation with the flux density at 25$\mu$m,
while LINER's and normal galaxies have a large spread. We also notice
that the bolometric flux of Seyferts and Starbursts have a large
contribution from emission in this waveband, as well as from other
infrared bands, indicating that the bolometric flux of these objects is
dominated by reradiation of visual and ultraviolet emission absorbed by
dust.

On the other hand, when we compare the bolometric flux with the flux
density at 7000\AA, we see that all galaxies present a good
correlation, but there is a clear separation between the normal
galaxies and the active ones (Seyfert 2's and Starbursts). The flux in
this waveband is the most important contributor to the bolometric flux of
normal galaxies, which are dominated by emission from the old stellar
population, but does not contribute significantly to the bolometric
flux of the more active objects.

\section{Summary}

In this contribution we reviewed some of the works on the spatially
integrated Spectral Energy Distributions of nearby galaxies.
This data have several applications for the study of nearby and distant
galaxies, like their stellar population, bolometric luminosities,
classification of distant objects and K-corrections, among others.

Some of the important points to be summarized here are that galaxies of
similar morphological, or activity class have similar SEDs. When
comparing the SEDs of Seyfert 2's with those of Starbursts, we see that
Seyfert 2's are redder in the optical and ultraviolet, but stronger in
the X-rays. The mid/far-IR emission of Seyfert 2's and Starbursts is
stronger than that of normal galaxies and LINER's.  This fact has a
strong influence in the galaxies bolometric flux. Seyfert 2's and
Starbursts bolometric flux is dominated by infrared emission from dust
reradiation, while in normal galaxies the bolometric flux is dominated
by the emission from the old stellar population.

\acknowledgements{
We would like to thank D. Calzetti for comments.  This work was
supported by NASA grants NAGW-3757, and AR-06389.01-94A.}

\vfill
\end{document}